\documentclass[reprint,nofootinbib,superscriptaddress,amsmath,amssymb,aps,prd]{revtex4-1}
\usepackage{amsmath,mathrsfs,amsbsy,color,graphicx,bm,amsthm,amsfonts}
\usepackage{bbm}
\usepackage{times}
\usepackage{units}
\usepackage{dcolumn}
\usepackage{graphicx}
\usepackage{epsfig}
\usepackage{epstopdf}
\usepackage[colorlinks,linkcolor=red,anchorcolor=green,citecolor=blue,CJKbookmarks=True]{hyperref}
\DeclareMathSymbol{\shortminus}{\mathbin}{AMSa}{"39}
\usepackage{mathrsfs}
\usepackage{braket}
\usepackage{amssymb}
\usepackage{txfonts}
\usepackage{float}
\usepackage{enumitem}
\usepackage{multirow}
\usepackage{orcidlink}

\newcommand{\meq}[1]{(\ref{#1})}

\begin{document}
\title{A Universal Framework for Horizon-Scale Tests of Gravity with Black Hole Shadows}

\author{Wentao Liu\orcidlink{0009-0008-9257-8155}}
\affiliation{Lanzhou Center for Theoretical Physics, Key Laboratory of Theoretical Physics of Gansu Province, 
Key Laboratory of Quantum Theory and Applications of MoE,
Gansu Provincial Research Center for Basic Disciplines of Quantum Physics, 
Lanzhou University, Lanzhou 730000, China}
\affiliation{Institute of Theoretical Physics $\&$ Research Center of Gravitation,
Lanzhou University, Lanzhou 730000, China}

\author{Yang Liu\orcidlink{0000-0003-2721-2559}}
\affiliation{Purple Mountain Observatory, Chinese Academy of Sciences, No. 10 Yuanhua Road, Nanjing 210023, China}

\author{Di Wu\orcidlink{0000-0002-2509-6729}}
\affiliation{School of Physics and Astronomy, China West Normal University, Nanchong, Sichuan 637002, China}

\author{Yu-Xiao Liu\orcidlink{0000-0002-4117-4176}}
\email[]{liuyx@lzu.edu.cn (Corresponding author)} 
\affiliation{Lanzhou Center for Theoretical Physics, Key Laboratory of Theoretical Physics of Gansu Province, 
Key Laboratory of Quantum Theory and Applications of MoE, Gansu Provincial Research Center for Basic Disciplines of Quantum Physics, 
Lanzhou University, Lanzhou 730000, China}
\affiliation{Institute of Theoretical Physics $\&$ Research Center of Gravitation,
Lanzhou University, Lanzhou 730000, China}

\begin{abstract}

In this Letter, we develop a numerically efficient framework for evaluating parameters in metric theories of gravity, and apply it to constrain the horizon-scale magnetic field in the Kerr–Bertotti–Robinson spacetime using the latest Event Horizon Telescope observations. 
The method’s adaptive ray-tracing strategy achieves near-linear computational efficiency without loss of numerical accuracy. 
This efficiency allows for high-precision black hole shadow modeling at a minimal computational cost and, for the first time, supports statistically robust parameter inference for arbitrary stationary black holes from horizon-scale observations.
Applying the framework to the exact magnetized and rotating Kerr–Bertotti–Robinson solution [\href{https://doi.org/10.1103/rfgv-ybz5}{Phys.~Rev.~Lett.~$\textbf{135}, 181401 (2025)$}], we evaluate the horizon-scale magnetic fields of M87* and Sgr A*. 
The analysis yields a field strength of $93.3^{+14.7}_{-23.8}\,\mathrm{G}$ for Sgr A*, consistent with the $71\,\mathrm{G}$ equipartition estimate from polarized ALMA observations, thereby supporting the predictions of Einstein's gravity.
\end{abstract}
\maketitle

\textit{Introduction.}\textemdash
In a major step forward, the  Event Horizon Telescope (EHT) Collaboration has recently released polarized images of M87* at 230 GHz, obtained during the 2017, 2018, and 2021 observing campaigns \cite{EventHorizonTelescope:2025vum}.
The inclusion of the 12-m Kitt Peak Telescope and the Northern Extended Millimetre Array in 2021 substantially improved the baseline coverage of the EHT, yielding a more precise measurement of the M87* ring diameter of $43.9\pm0.6~\mu{\rm as}$, compared to the $42\pm3~\mu{\rm as}$ value derived from the 2017 observations and published by the EHT Collaboration in 2019 \cite{EventHorizonTelescope:2019dse,EventHorizonTelescope:2019uob,EventHorizonTelescope:2019jan,EventHorizonTelescope:2019ths}.
These unprecedented observations have, in turn, motivated extensive theoretical and numerical efforts to interpret the ring morphology and polarization signatures \cite{EventHorizonTelescope:2020qrl,Cunha:2020azh,Crinquand:2022ogr,Davelaar:2021eoi,Cunha:2022gde,Galishnikova:2022mjg,Chen:2024nua,Uniyal:2025uvc}. 
In particular, the remarkable contributions from several groups \cite{Cunha:2018acu,Mizuno:2018lxz,Gralla:2019xty,Perlick:2021aok,Bronzwaer:2021lzo,Konoplya:2021slg,Vagnozzi:2022moj,Bambi:2019tjh,Chen:2022scf,Ayzenberg:2023hfw}, most notably the development of ray-tracing techniques \cite{Cunha:2015yba,Cunha:2019dwb,Hu:2020usx,Bacchini:2021fig,Zhang:2025xnl}, have enabled the imaging of non-integrable and fully numerical black hole spacetimes, providing a crucial bridge between theory and EHT observations.

However, high-resolution backward ray-tracing simulations are notoriously computationally intensive, making it difficult to apply statistical inference methods such as Markov Chain Monte Carlo (MCMC) directly to constrain model parameters from observational data \cite{Johannsen:2015hib,Palumbo:2022wnl,Yfantis:2024eqc,Chang:2025hrk}. 
To overcome this limitation, we construct a general framework that unifies numerical imaging and parameter estimation in metric theories of gravity.
Our approach employs an adaptive backward ray-tracing algorithm, inspired by the adaptive mesh refinement techniques in numerical relativity \cite{Clough:2015sqa,Buschmann:2021sdq}, which reduces the computational scaling from geometric to nearly linear.
This algorithm is combined with a Schwarzschild-calibrated normalization scheme that directly maps simulated image sizes to physical observables, independently of the observer’s configuration.
Consequently, our framework enables high-precision shadow modeling at a minimal computational cost and, for the first time, allows statistically robust inference of black hole parameters from horizon-scale observations for arbitrary black hole spacetimes.

In this Letter, we apply the above framework to the recently obtained Kerr–Bertotti–Robinson (Kerr–BR) black hole, an exact magnetized and rotating solution to the Einstein field equations \cite{Podolsky:2025tle}.  
Unlike the Kerr-Melvin spacetimes \cite{Ernst:1976bsr}, the Kerr–BR spacetime belongs to the Petrov type-D class and features finite, asymptotically uniform electromagnetic fields and bounded ergoregions, thereby providing a more realistic model for astrophysical black holes embedded in external magnetic environments \cite{Ovcharenko:2025cpm,Wang:2025bjf,Wang:2025vsx,Ali:2025beh,Zeng:2025olq,Zhang:2025ole}. 
By matching the theoretical predictions of the Kerr–BR spacetime with the latest EHT observations of M87* and Sgr A*, we place quantitative constraints on the magnetic field strength.
In particular, the analysis yields a magnetic field strength of $93.3^{+14.7}_{-23.8}\,\mathrm{G}$ for Sgr A*, consistent with the equipartition value $71\,\mathrm{G}$ inferred from polarized observations of submillimeter flares by the Atacama Large Millimeter/submillimeter Array (ALMA) \cite{Michail:2023mng}.
Our results suggest that the magnetized environment can be consistently described by the exact Kerr–BR solution within general relativity (GR), without invoking any modification to the theory.






\textit{Models.}\textemdash
\begin{figure*}[t]
\centering
\includegraphics[width=0.8\linewidth]{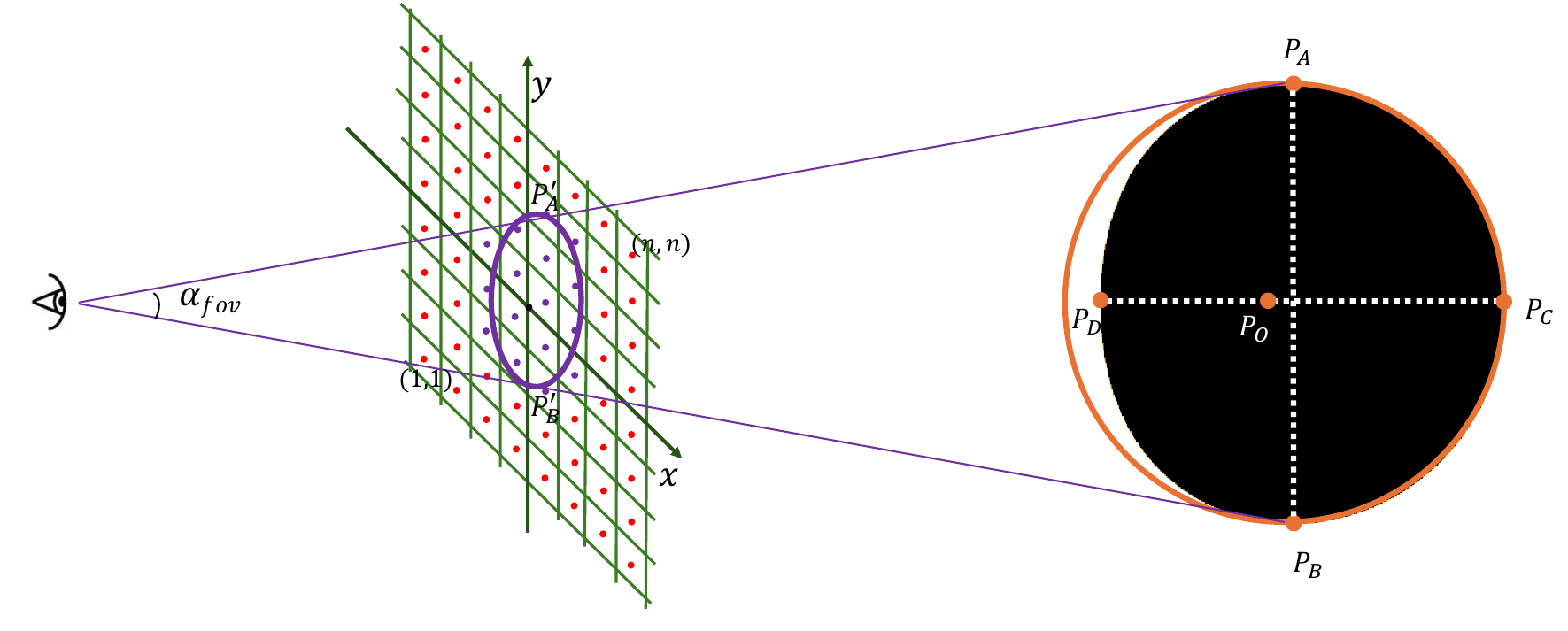}
\caption{
Schematic of the backward ray-tracing setup. 
A virtual observer at large distance defines the image plane $(x,y)$ with field of view $\alpha_{\rm fov}$.
Each grid point $(m,n)$ corresponds to a photon integrated backward in time until it crosses the horizon or escapes to infinity.
The purple curve is the \emph{apparent shadow contour} obtained from the captured/escaping photon separator.
The orange circle is a \emph{reference circle} used for noncircular shadows; it does not coincide with the shadow but serves as a comparator to quantify size and deformation (e.g., via characteristic points $P_A,P_B,P_C,P_D$).}
\label{fig1}
\end{figure*}
We start with the Kerr-BR black hole recently obtained in Ref. \cite{Podolsky:2025tle}, which represents an exact rotating black hole solution immersed in a uniform magnetic field within GR. 
The spacetime line element reads  
\begin{equation}
\begin{aligned}\label{ds2}
ds^2=\frac{1}{\Omega^2}&\bigg[-\frac{Q}{\Sigma}\big(dt-a\sin^2\vartheta d\varphi \big)^2+\frac{\Sigma}{Q}dr^2\\
&+\frac{\Sigma}{P}d\vartheta^2+\frac{P}{\Sigma}\sin^2\vartheta \big( adt-(r^2+a^2)d\varphi^2 \big)\bigg],
\end{aligned}
\end{equation}
where the metric functions take the form
\begin{equation}
\begin{aligned}
\Sigma&=r^2+a^2\cos^2\vartheta,\quad  P=1+B^2\big( M^2I_2/I_1^2-a^2 \big)\cos^2\vartheta,\\
Q&=(1+B^2r^2)\Delta,\quad ~\Omega^2=(1+B^2r^2)-B^2\Delta \cos^2\vartheta,\\
\Delta&=\big( 1-B^2M^2I_2/I_1^2 \big)r^2-2MrI_2/I_1+a^2,
\end{aligned}
\end{equation}
with $ I_1=1- B^2a^2/2 $ and $I_2=1- B^2a^2$. 
The metric reduces to the Kerr solution in the limit $B=0$, but not to the Schwarzschild spacetime for $a=0$, reflecting the intrinsic coupling between rotation and the external magnetic field. 
Here $M$ is an integration constant interpreted as the mass parameter of the black hole.
The outer and inner horizons of the black hole are determined by the roots of the metric function $\Delta$, namely $r_{\pm}\!=\!\tfrac{M I_1 I_2 \pm I_1\sqrt{M^2 I_2 - a^2 I_1^2}}{I_1^2 - B^2 M^2 I_2}$.
The explicit form of the electromagnetic field is given in Ref. \cite{Podolsky:2025tle}, but it is not essential to our present analysis; we are concerned only with its overall strength.

In this Letter, our goal is not to replace established analytical shadow methods \cite{Grenzebach:2014fha,Grenzebach:2015uva,Mars:2017jkk} that are widely used for analytically tractable spacetimes, which are valuable for shadow boundaries and parameter degeneracies.
Instead, our aim is to develop an efficient inference framework for extracting black hole parameters from shadow observations in more general, especially non-integrable, spacetimes.
To this end, we develop a new adaptive ray-tracing framework and use the Kerr-BR spacetime as a benchmark testbed, since its analytically tractable shadow allows a clean validation of the method \cite{Wang:2025vsx}. 
In the present Kerr–BR application, the framework directly links the observed shadow morphology to the strength of the underlying uniform magnetic field.
Photon trajectories are integrated using the Hamiltonian formulation of null geodesics in the Kerr–BR spacetime,
\begin{equation}\label{EqHJ}
\mathcal{H}=\tfrac{1}{2}g^{\mu\nu}p_{\mu}p_{\nu}=0,
\end{equation}
where $p_\mu=g_{\mu\nu}p^\nu$ and $p^\mu=(\dot{t},\dot{r},\dot{\vartheta},\dot{\varphi})$. 
The Killing vectors $\xi^\mu$ and $\psi^\mu$ yield two conserved quantities, the energy $\mathcal{E}$ and axial angular momentum $L_z$ \cite{Waldbook}, which allow us to numerically integrate the null geodesic equations in our backward ray-tracing scheme. 
Further computational details are given in Refs. \cite{Liu:2025lwj,Liu:2024lbi,Liu:2024lve,Liu:2024soc,Liu:2024iec}.
As illustrated in Fig. \ref{fig1}, a virtual observer is placed at a large radial distance, where the spacetime is effectively flat. 
The observer’s image plane, spanned by Cartesian coordinates $(x,y)$, is divided into an adaptive grid representing different initial photon directions within a field of view $\alpha_{\rm fov}$. 
Each grid point $(m,n)$ corresponds to a null geodesic that is integrated backward in time until the photon either crosses the event horizon or escapes to infinity. 
In our simulations, these two outcomes are distinguished on the image plane:  escaping photons are marked by red pixels, while captured photons are marked by purple pixels. 
After all grid points have been traced, the resulting collection of purple pixels forms the simulated black hole shadow, whose outer boundary is illustrated by the purple ring in Fig. \ref{fig1}.
For quantitative analysis, we introduce a reference circle, shown as the orange ring in Fig. \ref{fig1}, to compare with the noncircular shadow. 
This circle is defined as the circumcircle that passes through three characteristic points on the shadow boundary: $P_A(x_A,y_A)$\footnote{The normalized coordinates $(x_i, y_i)\in(0,1]$ are defined by the image resolution, e.g., $x_A = m/n$ for the $m$-th pixel in an $n\times n$ grid.}, $P_B(x_B,y_B)$, and $P_C(x_C,y_C)$. 
Meanwhile, a fourth point, $P_D(x_D, y_D)$, serves as a key indicator of the shadow's deformation.  
Denoting the circumcenter coordinates by $P_O(x_O,y_O)$, the center satisfies the following system of equations:
\begin{equation}
\begin{aligned}
(x_O - x_A)^2 + (y_O - y_A)^2 &= (x_O - x_B)^2 + (y_O - y_B)^2,\\
(x_O - x_A)^2 + (y_O - y_A)^2 &= (x_O - x_C)^2 + (y_O - y_C)^2.
\end{aligned}
\end{equation}
According to the Pythagorean theorem, the pixel coordinates $(x_i, y_i)$ can be used to determine the key geometric quantities of the shadow: 
the radius of the reference circle $R_s$ and the axial ratio $\mathcal{D}_s$ \cite{EventHorizonTelescope:2019dse}, which quantifies the deviation from circularity. 
They are given by
\begin{equation}
\begin{aligned}
\!\!  &R_s = \sqrt{(x_O - x_A)^2 + (y_O - y_A)^2},\\
\!\! &\mathcal{D}_s \!  =\! \sqrt{[(x_A \!-\! x_B)^2 + (y_A \!-\! y_B)^2]/[(x_C \!-\! x_D)^2 + (y_C \!-\! y_D)^2]}.
\end{aligned}
\end{equation}
Here, $R_s$ and $\mathcal{D}_s$ are observable quantities that are directly related to astronomical measurements, allowing one to constrain the theoretical model by fitting to observational data \cite{Hioki:2009na,Amarilla:2011fx}.

Through numerical backward ray-tracing simulations in the Kerr–BR spacetime, we establish explicit relationships linking observable quantities to black hole parameters and observer configuration.  
Accordingly, one can express $R_s \!\equiv\! R_s(\psi)$ and $\mathcal{D}_s \!\equiv\! \mathcal{D}_s(\psi)$ with $ \psi\!\equiv\! \{ \tilde{a},\tilde{B},\vartheta_0 \}$, where $\tilde{a}\!=\! a/M$ and $\tilde{B}\!=\!BM$ denote the dimensionless spin and magnetic field strength, respectively, and $\vartheta_0$ is the observer's inclination angle.
For each parameter set $(\tilde a,\tilde B,\vartheta_0)$, we ray-trace a two-dimensional shadow image on the observer’s image plane and extract the image-plane observables $(R_s,\mathcal{D}_s)$ from its pixels.
Because $R_s$ is an image-plane quantity, we convert it into an apparent angular radius, $\theta_s$, by applying a Schwarzschild-based calibration such that $\theta_s \propto R_s$.
Further details of this calibration are provided in Appendix \ref{SecA}.
With this calibration, the mappings $\theta_s(\psi)$ and $\mathcal{D}_s(\psi)$ provide a direct link between $\psi$ and the observable shadow features.
This link allows us to use EHT measurements to constrain the black hole parameters $\tilde{a}$, $\tilde{B}$ and $\vartheta_0$.
To achieve this, we construct the likelihood function as
\begin{align}\label{eq:likelihood}
\mathcal{L}(D_\mathrm{tot}|\psi)=\mathcal{P}_{\theta_s}(\theta_{\rm EHT}, \sigma_\theta|\psi) 
 \times  \mathcal{P}_{\mathcal{D}_s}(\mathcal{D}_{\rm EHT}, \sigma_{\mathcal{D}}|\psi),
\end{align}
when both observables $\theta_s$ and $\mathcal{D}_s$ are available.
Here, $D_\mathrm{tot}\!\equiv\!\{\theta_{\rm EHT}, \sigma_\theta, \mathcal{D}_{\rm EHT}, \sigma_{\mathcal{D}} \}$ denotes a set of EHT observational data.
In general, the probability densities $\mathcal{P}_{\theta_s}$ and $\mathcal{P}_{\mathcal{D}_s}$ are assumed to follow Gaussian distributions centered at the observed mean values $\theta_{\rm EHT}$ and $\mathcal{D}_{\rm EHT}$ with standard deviations $\sigma_\theta$ and $\sigma_{\mathcal{D}}$, respectively, when the reported uncertainties are symmetric.
For asymmetric uncertainties, they are modeled as  piecewise Gaussian distributions, with different standard deviations on each side of the mean. 
Then, the posterior distribution of theoretical parameters is obtained via Bayes' theorem: $ \mathcal{P}(\psi|D_\mathrm{tot})\propto \mathcal{L}(D_\mathrm{tot}|\psi)~\Pi(\psi)$, where $\Pi(\psi)$ represents the prior distribution, which we take to be uniform.
We employ the \textit{cobaya} MCMC sampler to explore this posterior distribution and obtain the constraints on parameters  $\tilde{a}$, $\tilde{B}$, and $\vartheta_0$.
Convergence is assessed using the Gelman-Rubin statistic, and the chains are regarded as converged once ($R\!-\!1\!<\!0.02$).

\begin{figure}[h]
\centering
\includegraphics[width=0.7\linewidth]{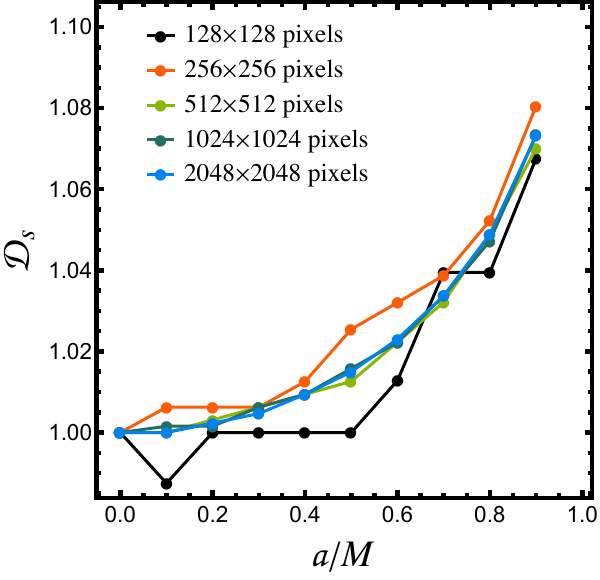}
\caption{Variation of the axial ratio $\mathcal{D}_s$ with the spin parameter $a/M$ for the Kerr black hole, extracted from images simulated at different pixel resolutions.}
\label{fig2}
\end{figure}

Our overall modeling strategy is conceptually straightforward, but its implementation involves several key technical challenges. 
The first concerns how to map the pixel coordinates of the characteristic points $P_{A,B,C,D}$ onto physically meaningful quantities relevant to astronomical observations, while the second relates to maintaining numerical precision and computational efficiency. 
The accuracy of the extracted shadow features is highly sensitive to the image resolution, as illustrated in Fig. \ref{fig2} for the Kerr case where the axial ratio $\mathcal{D}_s$ varies with spin.
The figure shows that numerical convergence is achieved only at sufficiently high resolutions and that the curves become progressively smoother.
To address this, we develope an adaptive ray-tracing algorithm that performs high-resolution calculations only in the vicinity of the critical pixels.
This adaptive scheme reduces the computational cost from exponential to nearly linear scaling with resolution, enabling us to explore the black hole parameter space at extremely high resolutions. 
The algorithmic details and validation tests of this method are presented in Appendix \ref{SecB}.

\textit{Results.}\textemdash
Having established a numerically consistent and efficient framework for shadow modeling, we next apply it to the EHT targets M87* and Sgr A*.
This allows us to link their observed shadow features to theoretical predictions within the Kerr–BR spacetime and to quantitatively infer the underlying magnetic field strength $\tilde{B}$.
To facilitate the comparison between observations and simulations, the relevant observational quantities and corresponding model parameters are listed in Table \ref{tab1}, with the latter expressed in geometrized units $(c\!\! =\!\! G \!\! =\!\!M\!\!=\!\!1)$.
\begin{table}[h]
\renewcommand{\arraystretch}{1.25}
\centering
\setlength\tabcolsep{1.3mm}{
\begin{tabular}{ccccccccc}
\hline\hline
\multirow{1}{*}{\quad} & \multirow{1}{*}{ M87*}&\multirow{1}{*}{ Sgr A* } 
\\
\hline
 Ring diameter $d $  & $ 43.9\pm0.6~\mu as $   &  $51.8\pm2.3~\mu as $ \\
 Axial ratio   & $ <4:3$   &- \\  
 Mass $\mathcal{M}$  & $ (6.5\pm0.7) \times 10^9 M_\odot $  &   $ (4.0\pm0.7) \times 10^6 M_\odot $    \\
 Distance $\mathcal{D}_O$ & $ (16.8\pm 0.8) {\rm~Mpc} $  & $ (8.3\pm 0.8) {\rm~kpc} $  \\
    $r_0$     &  $5.40\times10^{10}$   & $4.33\times 10^{10}$ \\
    $  \alpha_{\rm fov}  $     &  $1.25\times 10^{-10}\pi   $   & $1.56\times 10^{-10}\pi   $ \\
\hline\hline
\end{tabular}}
\caption{ Observed and simulated parameters used for the M87* and Sgr A* analyses. }
\label{tab1}
\end{table}
The ring diameters and axial ratios for M87* are taken from the EHT Collaboration’s 2025 and 2019 results \cite{EventHorizonTelescope:2025vum,EventHorizonTelescope:2019dse}, while those for Sgr A* are adopted from the 2022 EHT measurements \cite{EventHorizonTelescope:2022wkp}.  
The masses and distances follow the same observational estimates.  
The parameter $r_0$ represents the observer’s distance in the ray-tracing simulations, expressed in geometrized units according to the actual source–Earth separation.  
The field of view $\alpha_{\rm fov}$ is determined self-consistently by the calibration. 
Once these two parameters are fixed, the Schwarzschild shadow-area fraction $A_{\rm sh}$ is uniquely specified; see Appendix \ref{SecA} for details.

For the two EHT targets, M87* and Sgr A*, substituting the corresponding parameters from Table \ref{tab1} into the above expression yields the theoretical angular radii  $\theta_{s} $, which serve as the model predictions for subsequent parameter inference.
Before comparing these theoretical predictions with observations, it is essential to ensure the numerical convergence of the simulated shadow contours.
In general, the simulated shadow contours for different parameter choices converge at comparable resolutions, as illustrated in Fig. \ref{fig2}. 
Therefore, for calculating $\theta_s$ and $\mathcal{D}_s$, we find that an effective resolution of 4K (i.e., $4096\times4096$ pixels) is sufficient to ensure full numerical convergence.
With the numerical accuracy thus guaranteed, we proceed to constrain the model parameters using EHT observations.
To ensure sufficient sampling of the parameter space, we generate $10^4$ simulated shadow images for each of M87* and Sgr A* under randomly selected combinations of $(\tilde{a},\tilde{B},\vartheta_0)$.
We then interpolate this dataset to obtain continuous mappings of $(\theta_s,\mathcal{D}_s)$, which are used to calculate the posterior distributions of theoretical parameters.

\begin{figure}[t]
\centering
\includegraphics[width=1\linewidth]{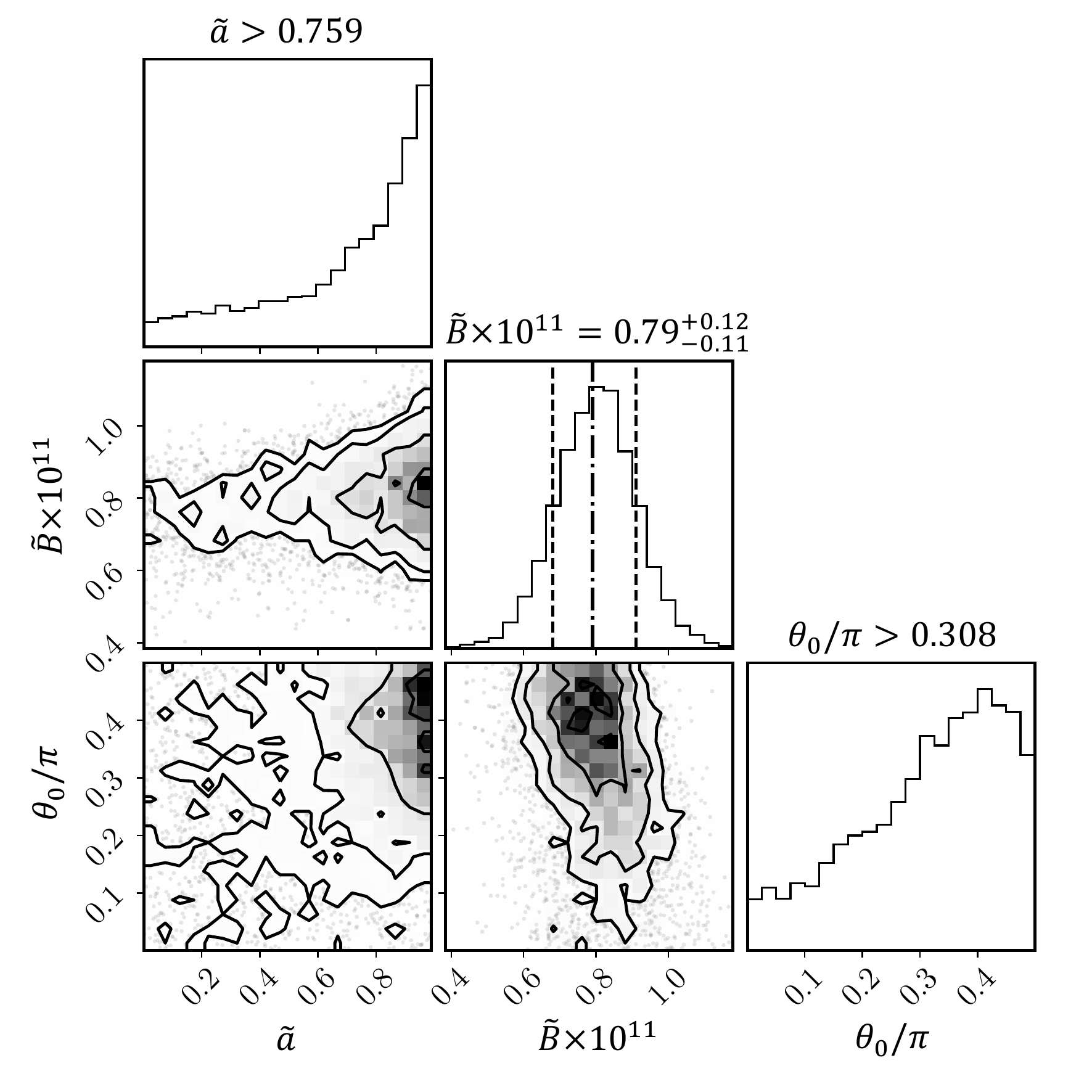}
\caption{Corner plot showing the one- and two-dimensional posterior probability distributions of the BH parameters for M87*, based on the three-epoch combined measurements from the EHT Collaboration \cite{EventHorizonTelescope:2025vum}.
The dashed lines indicate the $68\%$ CL, while the dot-dashed line marks the mean value of $\tilde{B}$.}
\label{fig3}
\end{figure}
\begin{figure}[t]
\centering
\includegraphics[width=1\linewidth]{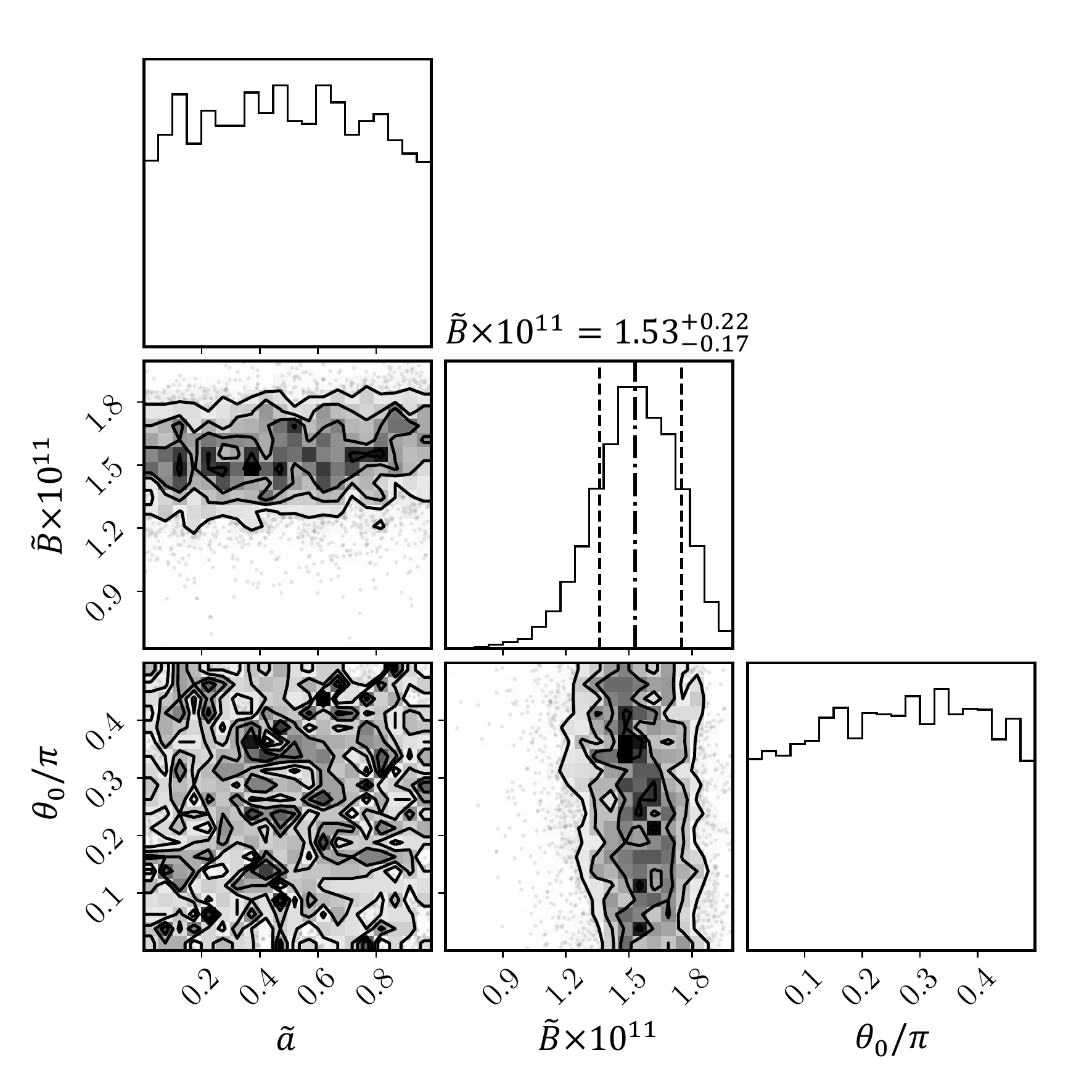}
\caption{
Corner plot showing the one- and two-dimensional posterior probability distributions of the BH parameters for Sgr A*, 
derived from the 2022 EHT observations \cite{EventHorizonTelescope:2022wkp} within the Kerr-BR spacetime. 
The dashed lines indicate the $68\%$ CL, while the dot-dashed line marks the mean value of $\tilde{B}$.}
\label{fig4}
\end{figure}
In Fig.~\ref{fig3}, we show the one- and two-dimensional posterior distributions of parameters for M87*, based on the three-epoch combined measurements of the EHT Collaboration \cite{EventHorizonTelescope:2025vum}.  
The magnetic field strength is constrained to $BM\!=\!0.79^{+0.12}_{-0.11}\times10^{-11}$.
The posterior peaks at a high spin\footnote{For the shadow observables considered here, the transformation \(\tilde a\to -\tilde a\) does not alter the relevant geometric features.}, $a/M \! >\! 0.759$, consistent with earlier jet-power arguments \cite{EventHorizonTelescope:2019ths}.
The inclination angle remains weakly constrained, with $\vartheta_0\!>\! \frac{77\pi}{250} $ at the 68\% confidence level (CL).  
Figure \ref{fig4} presents the corresponding posterior distributions for Sgr A*, derived using the EHT 2022 observational data \cite{EventHorizonTelescope:2022wkp}. 
In this case, the term $\mathcal{P}_{\mathcal{D}_s}$ in Eq. (\ref{eq:likelihood}) is omitted, since only $\theta_s$ is available (see Table \ref{tab1}).
The inferred magnetic field strength is higher, $BM \!=\! 1.53^{+0.22}_{-0.17}\times10^{-11} $, while the correlations among $a/M$, $BM$, and $\vartheta_0$ appear marginal.
This is not due to larger observational uncertainties but rather because  the Sgr A* dataset provides no effective constraint on the shadow Axial ratio $\mathcal{D}_s$, leaving one fewer observable to break parameter degeneracies.
Nevertheless, both analyses consistently favor the presence of a nonzero uniform magnetic field in the Kerr–BR spacetime, with the M87* results yielding a tighter credible interval for the field strength.

In this Letter, one of our main aims is to constrain the magnetic field strength by expressing the dimensionless product $BM$ in Gaussian units \cite{Hou:2022eev}, where the corresponding physical field is given by
\begin{equation}\label{BGauss}
B_{\rm Gauss} =  \frac{c^4}{G_{ N}^{3/2}M} (BM)\approx2.36\times10^{19} \left(\frac{M_\odot}{M}\right)(BM)\ {\rm G},
\end{equation}
where $B_{\rm Gauss}$ is expressed in gauss.
From the posterior distributions of $\tilde{B}$ and Eq. \meq{BGauss}, we derive  $B_{\rm Gauss}\!=\!0.0294^{+0.0045}_{-0.0055}\,{\rm G}$ (M87*)  and  $B_{\rm Gauss}\!=\!93.3^{+14.7}_{-23.8}\,{\rm G}$ (Sgr A*), both at 68\% CL.
The similar precision for Sgr A* and M87* despite their different ring-diameter uncertainties reflects the role of the black hole mass, source distance, and the central value of the ring diameter in shaping parameter inference.
Remarkably, the inferred field strength of Sgr A* is comparable to the equipartition magnetic field $(B_\text{Gauss}\!\!\sim\!\! 71\text{ G})$ derived from polarized ALMA observations of submillimeter flares \cite{Michail:2023mng}, \textcolor{black}{and is also of the same order as the $(B_\text{Gauss}\!\sim\! 29\text{ G})$ estimate obtained from the EHT one-zone model \cite{EventHorizonTelescope:2022urf}.
Meanwhile, the much weaker inferred field of M87* is qualitatively consistent with the expected decrease of horizon-scale magnetic field strength with black hole mass. 
Because our Kerr-BR framework constrains an effective magnetic-field parameter entering the spacetime description, rather than directly reconstructing the localized plasma field, and may capture only part of the overall magnetic environment, astrophysical estimates such as the 1-30G value for M87* \cite{EventHorizonTelescope:2021srq} are included only for order-of-magnitude comparison. 
Taken together, these comparisons provide a nontrivial consistency check that the Kerr-BR solution, as an exact magnetized black hole spacetime in GR, can yield field strengths broadly compatible in scale with independent polarimetric estimates when constrained by shadow geometry and EHT priors, even across systems differing in mass by more than three orders of magnitude.
This in turn suggests that horizon-scale magnetism may be captured, at least in part, within a purely GR-based geometric framework without invoking modifications of Einstein’s theory.}

\begin{table}[h]
\renewcommand{\arraystretch}{1.25}
\centering
\setlength\tabcolsep{2.5mm}
{\begin{tabular}{lccccc}
\hline\hline
Epoch  &  Ref. & $d$ [$\mu$as] &  $B_{\rm Gauss}$[G] \\
\hline
2017 (Sgr A* )       &2022 \cite{EventHorizonTelescope:2022wkp}   & $51.8^{+2.3}_{-2.3}$ &   $93.3^{+14.7}_{-23.8}$               \\
2018 (M87* )        &~2024 \cite{EventHorizonTelescope:2024dhe} & $43.3^{+1.5}_{-3.1}$  &   $0.0204^{+0.0103}_{-0.0114}$    \\
2017-2021(M87*)&2025 \cite{EventHorizonTelescope:2025vum}   & $43.9^{+0.6}_{-0.6}$ &   $0.0294^{+0.0045}_{-0.0055} $  \\
2017 (M87*)         &2025 \cite{EventHorizonTelescope:2025vum}   & [42.0, 46.4]                  &   $0.0285^{+0.0114}_{-0.0101}$    \\
2018 (M87*)         & 2025 \cite{EventHorizonTelescope:2025vum}  & [40.7, 44.4]                  &   $0.0216^{+0.0096}_{-0.0100}$     \\
2021 (M87*)         & 2025 \cite{EventHorizonTelescope:2025vum}  & [43.1, 44.5]                   &   $0.0290^{+0.0046}_{-0.0059}$      \\
\hline\hline
\end{tabular}}
\caption{Inferred magnetic field strengths $B_{\rm Gauss}$ of M87* and Sgr A* at the 68\%CL, derived from EHT observations across different epochs.}
\label{tab2}
\end{table}
Finally, we summarize all inferred constraints in Table \ref{tab2}.
In addition to the results presented in Figs. \ref{fig3} and \ref{fig4}, we incorporate other EHT measurements of M87* for comparison.
These include the 2018 data reanalysis published in 2024 \cite{EventHorizonTelescope:2024dhe} and the most recent three-epoch joint analysis that combines observations from 2017, 2018, and 2021 \cite{EventHorizonTelescope:2025vum}.
The latter study, which incorporated the 12-m Kitt Peak Telescope and the Northern Extended Millimetre Array, provided substantially improved baseline coverage and confirmed the long-term stability of the horizon-scale ring across all epochs.
Taken together, these results demonstrate that the Kerr-BR model yields magnetic field strengths consistent with the latest EHT constraints while remaining fully compatible with GR, thereby reinforcing the viability of horizon-scale magnetism as a purely geometric observable.


\textit{Remarks.}\textemdash
In summary, we have constructed a general and self-consistent framework that provides an efficient route for evaluating parameters in metric theories of gravity through black hole shadow geometry. 
Applying this framework to the Kerr–BR black hole spacetime \cite{Podolsky:2025tle}, we obtained constraints on the magnetic-field strengths of M87* and Sgr A*. 
The result for Sgr A* aligns with independent polarization estimates \cite{Michail:2023mng}, demonstrating that the observed horizon-scale magnetization can be naturally accommodated within Einstein’s gravity. 
Since the Kerr–BR black hole is an exact Einstein–Maxwell solution, however, the magnetic field inferred here should be interpreted not as the environmental magnetic field of M87* or Sgr A*, but as an effective horizon-scale magnetic field encoded in the Kerr–BR metric.
Because the actual near-horizon magnetic structure may naturally receive contributions from both this metric-encoded field and the surrounding astrophysical environment, the order-of-magnitude consistency with independent astrophysical estimates is physically reasonable and informative.

The framework developed here therefore offers a new geometric approach to probing near-horizon phenomena and shows that magnetized black hole configurations remain compatible with astrophysical observations within GR. 
It also provides a natural foundation for extending such analyses to more general theories of gravity. 
Beyond GR, theories involving additional dynamical fields or couplings introduce a much larger parameter space. 
The adaptive, inference-oriented design of our method, which eliminates the need for exhaustive parameter sweeps, makes it particularly well suited to these high-complexity models and to systematic observational tests of their signatures. 
By linking enlarged theoretical parameter spaces to measurable image features, this framework opens new opportunities for testing GR and its possible extensions with upcoming EHT and next-generation VLBI arrays.

\textit{Acknowledgments.}\textemdash
This work is supported in part by the National Natural Science Foundation of China (Grants No. 12475056, No. 12247101, No. 12205243), Gansu Province’s Top Leading Talent Support Plan, the Natural Science Foundation of Gansu Province (No. 22JR5RA389), and the 111 Project (Grant No. B20063).

%

\appendix
\section{Schwarzschild calibration}\label{SecA}
This Appendix provides the Schwarzschild calibration scheme employed in the main analysis.
For the mapping between pixel coordinates and physically measurable quantities, we establish a normalization procedure based on the Schwarzschild black hole. 
\subsection{Geometric normalization}
For a Schwarzschild black hole, the shadow size is absolute and depends only on its mass, independent of the observer’s location or orientation. 
After setting $M\!=\!1$, the physical shadow radius is fixed at $ R_{\rm phys}\!=\!3\sqrt{3}$. 
In numerical ray-tracing simulations, however, the apparent size of the shadow on the image plane depends on the observer’s setup, such as the field of view $\alpha_{\rm fov}$ and distance $r_0$ rather than being determined solely by the intrinsic physical scale.
We therefore use the Schwarzschild case ($\tilde{a}\!=\!\tilde{B}\!=\!0$) as a reference standard: by comparing the simulated image size of the Kerr-BR black hole with that of the Schwarzschild one under identical conditions, we can calibrate the apparent shadow scale on the image plane. 
To simulate the imaging relationship between Earth and the supermassive black holes M87* and Sgr A*, we take the actual source–observer distance as the parameter $r_0$. 
By appropriately adjusting the field of view $\alpha_{\rm fov}$, the simulated Schwarzschild shadows at different distances $r_0$ can be made to appear identical in size, effectively serving as a “standard candle” for image normalization.
This calibration allows us to disentangle the coupling between the magnetic field strength $\tilde{B}$ and the observational distance when analyzing the Kerr-BR black hole shadows. 

\begin{figure}[h]
\centering
\includegraphics[width=1\linewidth]{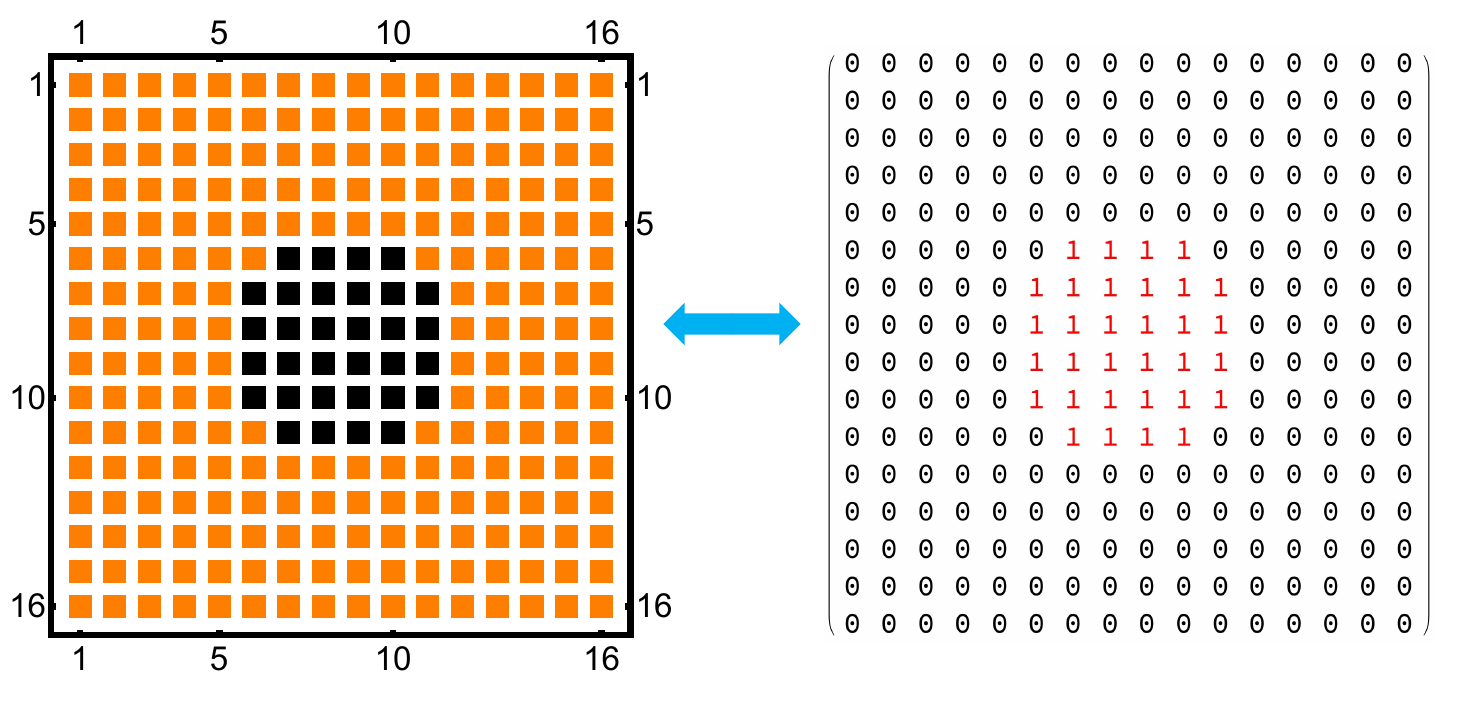}
\caption{Schematic illustration of the shadow-area fraction $ A_{\rm sh} $ for the Schwarzschild black hole.}
\label{figApp1}
\end{figure}
Before establishing the quantitative correspondence between $r_0$ and $\alpha_{\rm fov}$, we introduce the Schwarzschild shadow-area fraction $A_{\rm sh}$, defined as the ratio of the shadowed region to the total image-plane area, as illustrated in Fig. \ref{figApp1}.
The left panel shows a schematic example of a discretized image plane (here a $16 \times 16$ grid for illustration), in which black pixels denote the shadow region and orange pixels denote light rays reaching the observer.
The right panel shows the corresponding two-dimensional binary matrix, where 1 is assigned to shadow pixels and 0 otherwise.
The shadow-area fraction is then computed as the normalized sum of the matrix elements, which converges to the physical area fraction in the limit of sufficiently high-resolution.
Now, we perform a parameter sweep over $(r_0,\alpha_{\rm fov})$ in the Schwarzschild case and identify the combinations that yield identical pixel-scale shadow diameters, corresponding to the same Schwarzschild shadow-area fraction $ A_{\rm sh} $, in the simulated data.
\begin{figure}[h]
\centering
\includegraphics[width=1\linewidth]{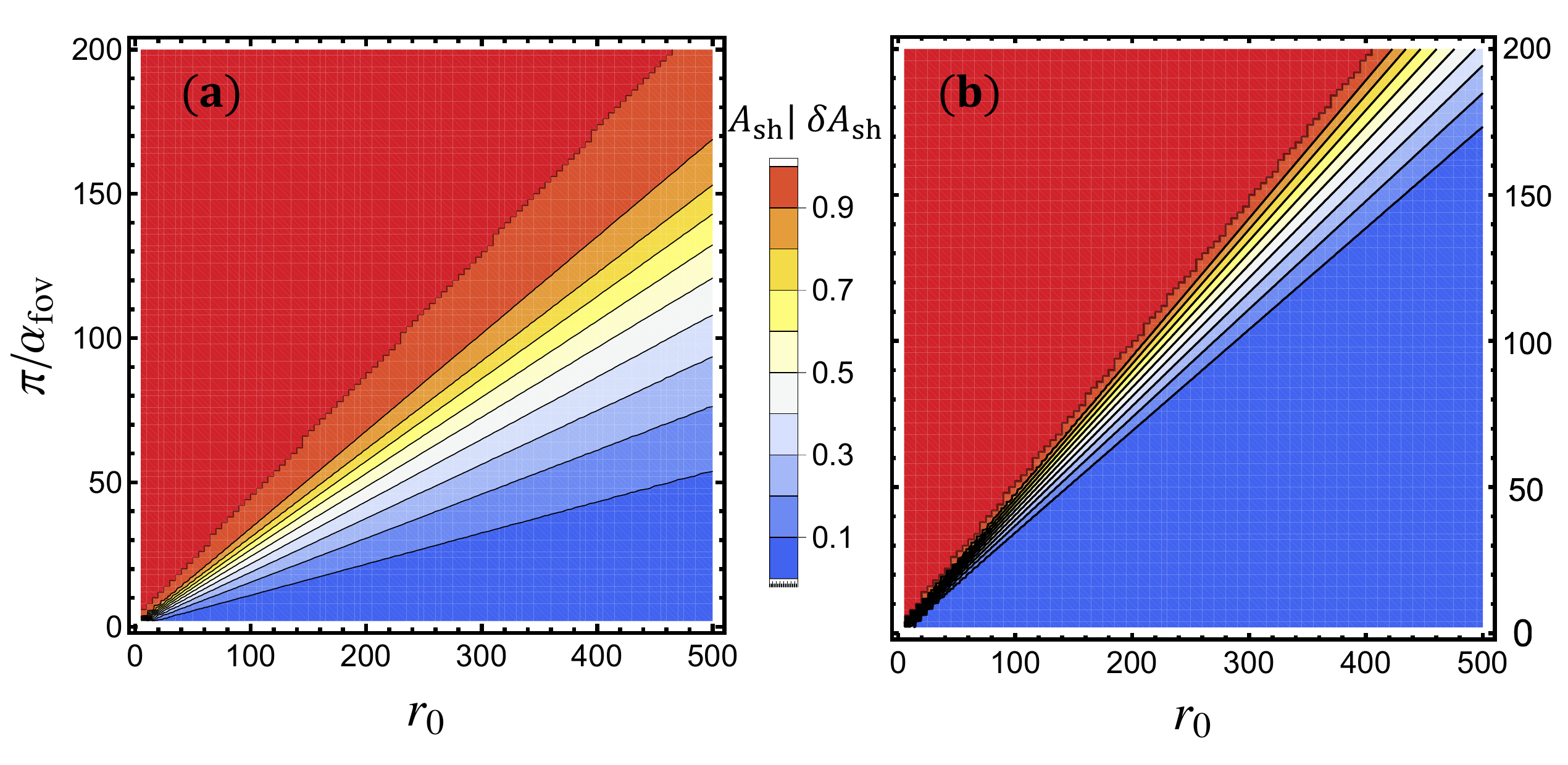}
\caption{(a) Contour map of the Schwarzschild shadow-area fraction $ A_{\rm sh} $. Contours show $(r_0,\,\pi/\alpha_{\rm fov})$ combinations that produce identical apparent shadow sizes, with the contour level representing the fractional shadow area (0–1 scale). 
(b) Area-fraction difference $\delta A_{\rm sh}$ between the empirical fit in Eq. \meq{Figeq} and the numerical data.}
\label{figApp2}
\end{figure}
The results of this parameter sweep are shown in Fig. \ref{figApp2}(a), where the mapping is expressed in terms of the inverse field of view parameter $\pi/\alpha_{\rm fov}$. 
Because $\alpha_{\rm fov}\in(0,\pi)$, the transformed variable spans $(1,\infty)$, providing a more linear and visually uniform representation of the $r_0$--$\alpha_{\rm fov}$ relation.
Each contour represents a set of $(r_0,\,\pi/\alpha_{\rm fov})$ pairs that produce simulated Schwarzschild shadows of identical apparent size on the image plane. 
The contour level denotes the Schwarzschild shadow-area fraction $A_{\rm sh}$, which varies from 0 to 1.
Contour levels exceeding $\pi/4\!\approx\!0.785$ are not physically meaningful, as they correspond to configurations where the shadow extends beyond the image boundaries.
As expected, larger observer distances correspond to larger values of $\pi/\alpha_{\rm fov}$, reflecting the need for smaller field of view angles to preserve the same apparent shadow diameter. 
This monotonic correspondence provides a quantitative calibration of the imaging geometry and allows us to express the mapping in analytical form. 
To facilitate subsequent applications, we model the shadow-area fraction with the empirical power law
\begin{equation}\label{Figeq}
A_{\rm sh}(r_0,\alpha_{\rm fov})=8.482 \left(\frac{\pi}{r_0~\alpha_{\rm fov} }\right)^2.
\end{equation}
Among the candidate fitting forms we tested, this quadratic power-law form provides the best overall accuracy and stability in the range $r_0\!>\!100$ relevant to our calibration and in the physically relevant regime of $A_{\rm sh}$.
As shown in Fig. \ref{figApp2}(b), the area-fraction difference $\delta A_\text{sh}$ between the empirical fit and the numerical data remains extremely small in this regime, thereby justifying its use as an analytic calibration for subsequent normalization and scaling.

Using the shadow-area fraction $A_{\rm sh}$, we obtain the proportional relation
\begin{equation}
A_{\rm sh}:\pi R_s^2\,=\,\pi\, \big(3\sqrt{3}\big)^2: \pi\, R_{\rm phys}^2,
\end{equation}
where equating the corresponding area ratios yields the mapping between the Schwarzschild-calibrated physical shadow radius $R_{\rm phys}$ and the image-plane radius $R_s$.

\subsection{Conversion to celestial angular radius}
With the Schwarzschild-calibrated physical shadow radius $R_{\rm phys}$ obtained,  the final step is to convert this calibrated shadow radius into the celestial angular radius $\theta_s$ observable from Earth. 
To this end, we employ the standard relation between the physical shadow radius and its apparent angular size, given by $ \theta_{s} \!=\! R_{\rm phys}\frac{\mathcal{M}}{D_O}$ \cite{Amarilla:2011fx}.
Based on our Schwarzschild-calibrated normalization scheme, the corresponding angular radius can be written in microarcseconds as
\begin{equation}
\theta_s = 9.87098 \times 10^{-6} R_s\frac{3\sqrt{3}}{\sqrt{A_{\rm sh}/\pi}} \left(\frac{\mathcal{M}}{M_\odot}\right) \left(\frac{1 \text{kpc}}{D_O}\right) \mu\text{as},
\end{equation}
where $R_s$ denotes, as previously discussed, the radius on the image plane.
It depends on the parameters $(r_0,\alpha_{\rm fov},\tilde{a},\tilde{B},\vartheta_0)$, and thus $\theta_{s}$ is also a function of these quantities.

\subsection{Consistency tests}

To assess the robustness of this calibration scheme, we perform consistency tests using the Kerr–BR spacetime as a representative example.
In recognition of the important contribution by Guo \emph{et al.} to the study of black hole shadows in this spacetime \cite{Wang:2025vsx},  we adopt the same observer configuration as used in their analysis—specifically, the setup shown in Fig. 1 of Ref. \cite{Wang:2025vsx},  with the observer located at $r_0\!=\!300$ and the field of view fixed at $\alpha_{\rm fov}\!=\!\pi/16$. 

Under these conditions, the apparent shadow of the Schwarzschild black hole occupies a fractional area of $  A_{\rm sh}\!=\!0.02413\!\simeq\! \pi/130$ on the image plane. 
We take this value as a “standard candle” to calibrate the shadow size in the Kerr–BR spacetime.
\begin{figure}[h]
\centering
\includegraphics[width=0.8\linewidth]{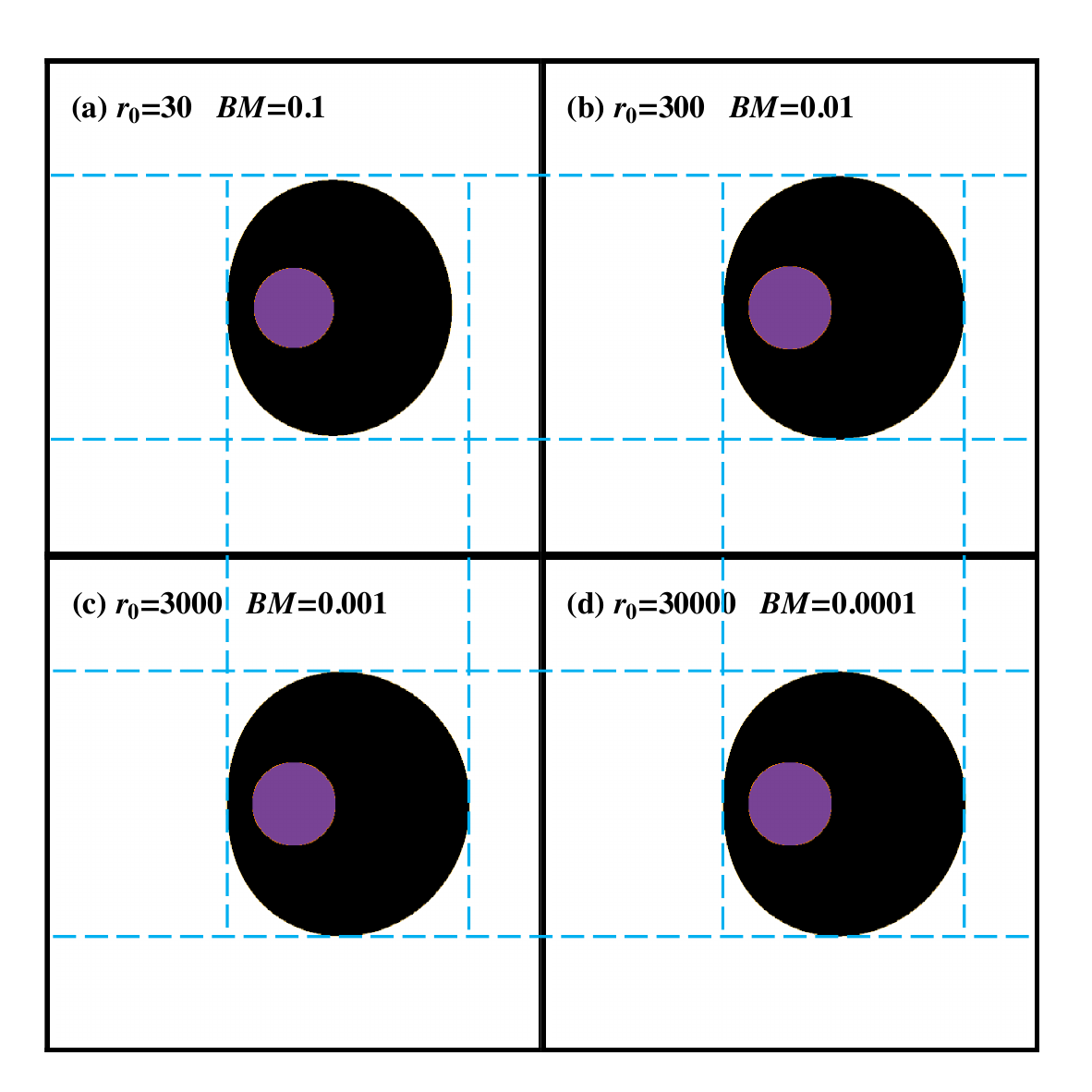}
\caption{Variation of the Kerr–BR black hole shadow with $\tilde{a}=0.94$ and $\vartheta_0=\frac{\pi}{2}$ under different values of $r_0$ and $\tilde{B}$, at a fixed $A_{\rm sh}=\pi/130$.}
\label{figApp3}
\end{figure}
In Fig. \ref{figApp3}, we present the simulated Kerr–BR black hole shadows at different observer distances $r_0$. 
The field-of-view angle $\alpha_{\rm fov}$ is adjusted according to Eq. \eqref{Figeq} so that the reference Schwarzschild shadow (purple circle) maintains the same apparent size in all panels, serving as a “standard candle” for direct comparison of magnetic field effects.
Although not central to our study, the Kerr–BR image in Fig. \ref{figApp3}(b) exactly matches Fig. 1 of Ref. \cite{Wang:2025vsx}, confirming the consistency of our setup. 
As indicated by the blue dashed guides, Figs. \ref{figApp3}(b)–\ref{figApp3}(d) exhibit nearly identical shadows, with larger observer distances requiring weaker magnetic fields to reproduce the same apparent scale.  
This demonstrates a distance–field-strength coupling that our calibration scheme removes.
This confirmation allows us to confidently apply the calibrated relation in Eq. \eqref{Figeq} to constrain the parameters of the Kerr–BR spacetime in the following analysis.

\section{Adaptive ray-tracing algorithm}\label{SecB}
This Appendix provides the details of the adaptive ray-tracing algorithm used to construct high-resolution black hole shadow images with controlled numerical accuracy.
As illustrated in Fig. \ref{fig2}, the accuracy of the extracted shadow observables, such as $R_s$ and $\mathcal{D}_s$, is highly sensitive to the image resolution.
However, the computational cost increases almost geometrically with higher resolution, making it impractical to generate the large ensemble of images required for the MCMC analysis. 
To address this issue, we develop an adaptive backward ray-tracing algorithm that dynamically refines the sampling resolution around the shadow boundary.

\begin{figure}[h]
\centering
\includegraphics[width=0.98\linewidth]{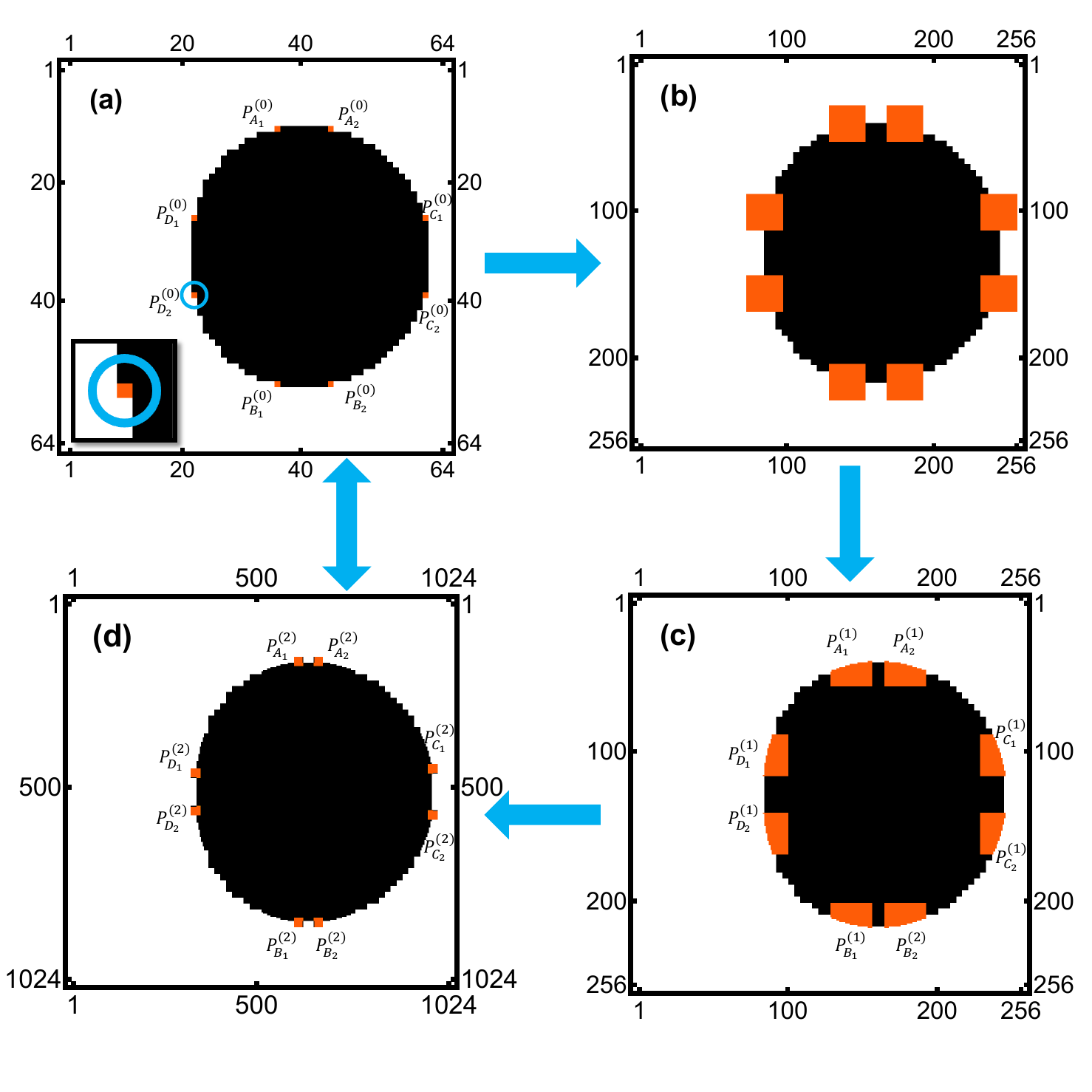}
\caption{Iterative “resolution-shooting” process of the adaptive backward ray-tracing algorithm.  
The base image (a) is progressively refined around the marked boundary regions, yielding high-resolution locating pixels after successive iterations.}
\label{figApp4}
\end{figure}
\begin{figure}[h]
\centering
\includegraphics[width=0.98\linewidth]{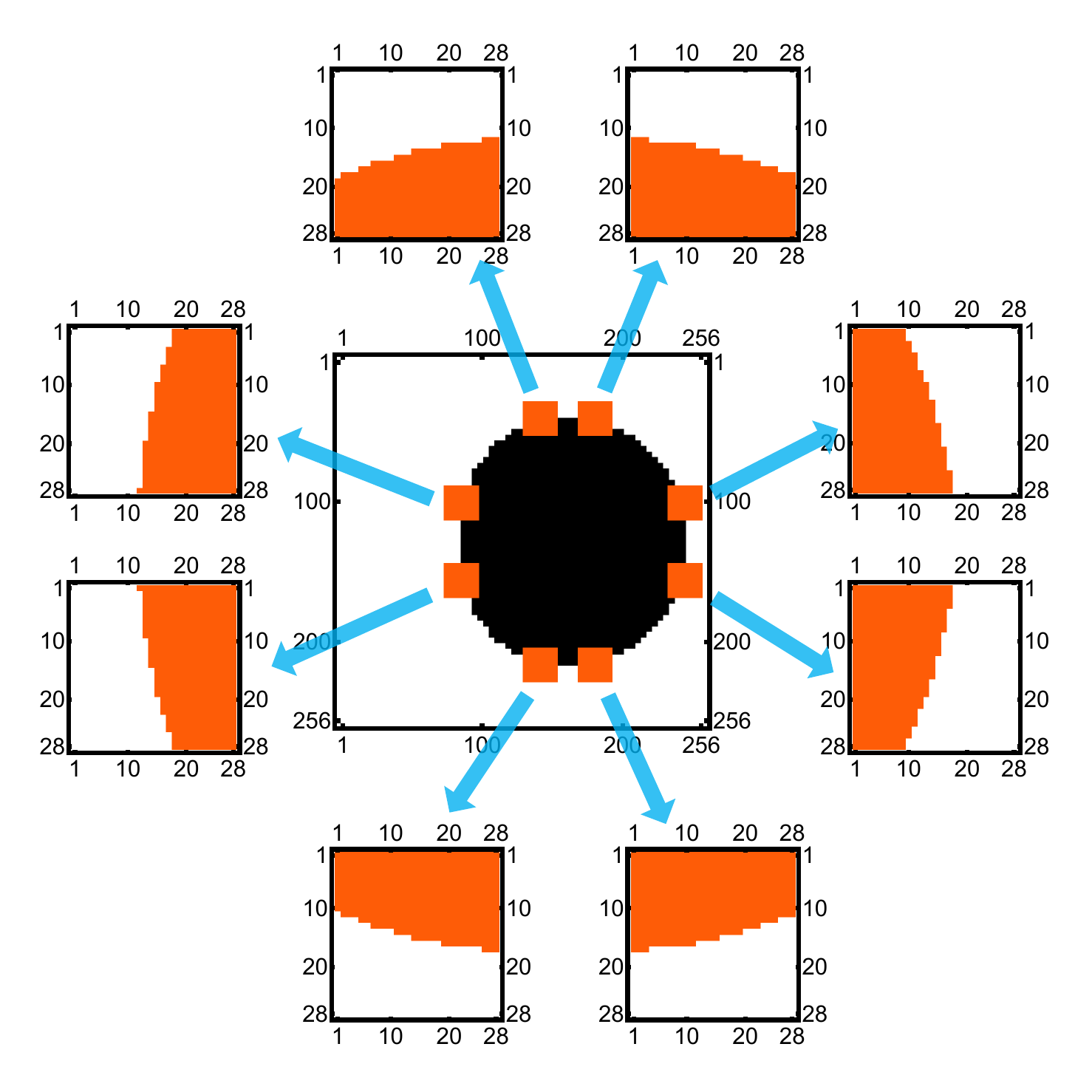}
\caption{The eight boundary patches surrounding the shadow edge are refined independently.  
Each refined subregion is then inserted back into the global image, providing updated locating pixels for the next iteration.} 
\label{figApp5}
\end{figure}
As shown in Fig. \ref{fig1}, the information required to determine the shadow observables depends solely on the coordinate accuracy of the characteristic points $P_i$, implying that a uniformly high image resolution outside these regions is unnecessary. 
Inspired by adaptive mesh refinement techniques used in numerical relativity, we employ a coarse-to-fine “resolution-shooting” iteration that begins with a coarse image and progressively refines only the regions relevant to the shadow boundary.
In our procedure, the target is to locate the characteristic points $P_i$, but they cannot be identified directly on a discrete image grid. 
Instead, for each $P_i$ we first bracket it by two locating endpoints $P_{i_1}$ and $P_{i_2}$ lying on opposite sides of the shadow edge. 
These endpoints serve as the operational handles of the algorithm, marking the pixel regions that require refinement and defining the characteristic point as $P_i \!\equiv\! \tfrac{1}{2}\,(P_{i_1}+P_{i_2})$.
This midpoint-based definition ensures that the characteristic points systematically converge toward the true boundary location as the resolution increases.
We start with a base image of $64\times64$ pixels, as shown in Fig. \ref{figApp4}(a), where the initial locating pixels at positions $P^{(0)}_{i_1}$ and $P^{(0)}_{i_2}$ are marked in orange to identify the boundary regions for subsequent refinement.
The base image is treated as a $64\times64$ matrix with pixel values 0 (white), 1 (black), and 2 (orange for marked regions). 
Each orange pixel is expanded to cover a local $(\pm n_1)$ neighborhood (controlling the size of the coarse boundary patch), forming eight $7\times7$ orange subregions for $n_1\!=\!3$. 
Subsequently, all pixels in the image are expanded into $n_2\times n_2$ submatrices (controlling the local refinement factor) and flattened to generate a super-resolved image with an effective resolution of order $(64n_2)\times(64n_2)$, as shown in Fig. \ref{figApp4}(b), where $n_2\!=\!4$.
At this stage, each marked local neighborhood contains $(n_1 n_2)\times(n_1 n_2) \!=\! 28\times28$ pixels. 
We then initiate the first iteration, performing backward ray-tracing calculations only within these eight $28\times28$ local neighborhoods, as illustrated in Fig. \ref{figApp5}.
The results of the eight refined subregions are inserted back into the original image, producing the updated map in Fig. \ref{figApp4}(c). 
This image accurately reproduces the characteristic point information that would be obtained from a full $256\times256$ ray-tracing simulation, with the updated locating pixels markers denoted as $P^{(1)}_{i_1}$ and $P^{(1)}_{i_2}$.
The procedure is then repeated for a second iteration, raising the local resolution to $1024\times1024$ and yielding $P^{(2)}_{i_1}$ and $P^{(2)}_{i_2}$.
Within this framework, the iterative refinement allows us to achieve sufficiently high effective resolution to accurately determine the locating pixels $P^{(N)}_j$ $(j\!=\!1,\dots,8)$,  where $N$ denotes the iteration number and the set $\{P^{(N)}_j\}\!=\!\{P^{(N)}_{A_1},\dots,P^{(N)}_{D_2}\}$.
By appropriately choosing $n_1$ and $n_2$, the algorithm reproduces the same locating pixels as a full all-pixel simulation while requiring only a fraction of its computational resources.

Moreover, an internal self-consistency check is introduced by defining a standard-deviation criterion, denoted as
\begin{equation}
\begin{aligned}
&\!\!\!\sigma^{(N)}\! =\! \sqrt{\frac{1}{16\!-\!1}\sum^{16}_{k=1} 
 \bigg[ E^{(N)}_k\!-\!\bar{E}^{(N)}  \bigg]^2  },\!
\quad \bar{E}^{(N)}=\frac{1}{16}\sum^{16}_{k=1} E^{(N)}_k,\\
&\!\!\!E^{(N)}_{j,x}=100\frac{x^{(N)}_j-x^{(N-1)}_j}{x^{(N-1)}_j}, \quad \!
E^{(N)}_{j,y}=100\frac{y^{(N)}_j-y^{(N-1)}_j}{y^{(N-1)}_j},
\end{aligned}
\end{equation}
where $E^{(N)}_k\!=\!\{E^{(N)}_{1,x},\dots,E^{(N)}_{8,y}\}$ denotes the set of percentage corrections for all coordinate components in the current iteration, computed from the eight locating pixels $P^{(N)}_j$ with coordinates $\{x^{(N)}_j, y^{(N)}_j\}$.
The coordinate differences of the locating pixels (i.e., the changes in their $x$- and $y$-positions) are expressed as relative percentages with respect to the previous iteration. 
This allows $\sigma^{(N)}$ to serve as a dimensionless indicator of numerical convergence, quantifying the stability of the locating pixels across iterations and acting as an internal consistency measure.
As the iteration proceeds, $\sigma^{(N)}$ is expected to decrease toward zero, with convergence achieved when $\sigma^{(N)} \!<\! \varepsilon$. 
If $\sigma^{(N)}\! >\! \sigma^{(N-1)}$ and $\bar{E}^{(N)} \!>\! \bar{E}^{(N-1)}$, the algorithm dynamically adjusts by increasing $n_1$ and decreasing $n_2$. 
This adjustment widens the coarse boundary patch while reducing the refinement level until a reliable search region is re-established.
Such an error-controlled feedback mechanism makes the method inherently adaptive.
Ultimately, the adaptive ray-tracing framework developed here provides both accuracy and efficiency for shadow modeling.

\begin{figure}[h]
\centering
\includegraphics[width=0.98\linewidth]{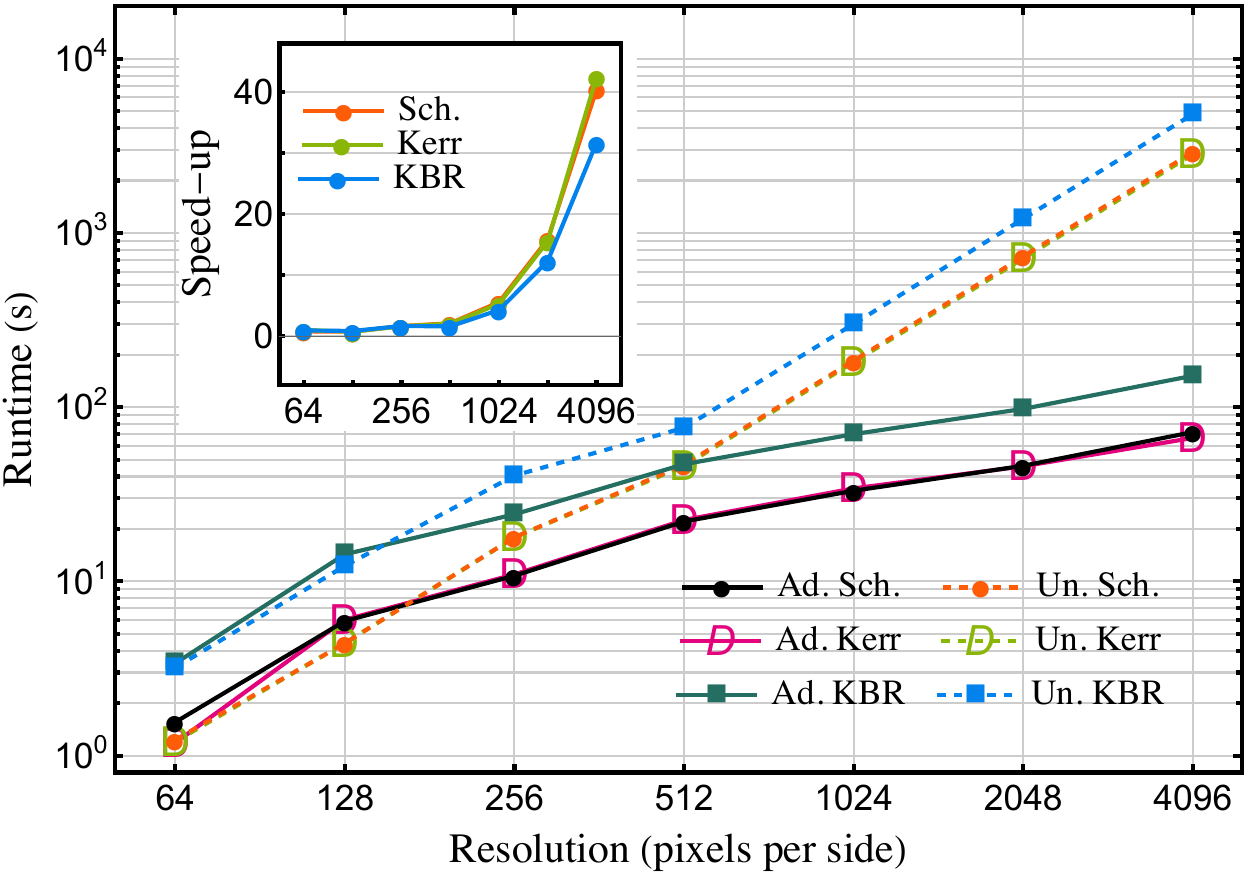}
\caption{Runtime benchmark of the adaptive and uniform ray-tracing schemes for Schwarzschild, Kerr, and Kerr–BR black holes. The inset shows the speed-up factor $T_\text{Uniform}/T_\text{Adaptive}$.} 
\label{figApp6}
\end{figure}
Finally, we perform runtime benchmarks comparing the adaptive ray-tracing algorithm with the corresponding uniform full-image calculations. 
All calculations are carried out on an \textit{AMD Ryzen~9~9950X} platform using \textit{Wolfram Mathematica}, and the same observational setup as in Fig.~\ref{figApp3}(b) is adopted, so as to assess the computational efficiency and scalability of the adaptive ray-tracing algorithm under a representative, uniform numerical setup. 
Figure~\ref{figApp6} compares the runtime of the adaptive and uniform full-image calculations over a range of image resolutions for three representative black-hole configurations: Schwarzschild \((\tilde{a}=0,\tilde{B}=0)\), Kerr \((\tilde{a}=0.94,\tilde{B}=0)\), and Kerr-BR \((\tilde{a}=0.94,\tilde{B}=0.01)\). 
One can see that, although the adaptive scheme may take slightly longer than the corresponding uniform calculation at very low resolutions, its runtime growth is strongly suppressed as the target resolution increases, eventually leading to a substantial computational advantage. 
The inset in Fig.~\ref{figApp6} further shows the corresponding speed-up factor, defined as the ratio of the runtime of the uniform calculation to that of the adaptive one. 
As the resolution increases, this speed-up gain also grows systematically and rapidly. 
These benchmarks demonstrate the strong computational advantage and scalability of our adaptive method.

\end{document}